# A mutual information-based *in vivo* monitoring of adaptive response to targeted therapies in melanoma

Aurore Bugi-Marteyn[1,2†], Fanny Noulet[1,2†], Nicolas Liaudet[3†] and Rastine Merat[1,2*]

[1]*Dermato-Oncology Unit, Division of Dermatology, University Hospital of Geneva, Switzerland*
[2]*Department of Pathology and Immunology, Faculty of Medicine, University of Geneva, Switzerland*
[3]*Bioimaging core Facility, Faculty of Medicine, University of Geneva, Switzerland*
(Dated: December 7, 2020)

The mechanisms of adaptive resistance to genetic-based targeted therapies of solid malignancies have been the subject of intense research. These studies hold great promise for finding co-targetable hub/pathways which in turn would control the downstream non-genetic mechanisms of adaptive resistance. Many such mechanisms have been described in the paradigmatic *BRAF*-mutated melanoma model of adaptive response to BRAF inhibition. Currently, a major challenge for these mechanistic studies is to confirm *in vivo*, at the single-cell proteomic level, the existence of dependencies between the co-targeted hub/pathways and their downstream effectors. Moreover, the drug-induced *in vivo* modulation of these dependencies needs to be demonstrated. Here, we implement such single-cell-based *in vivo* expression dependency quantification using immunohistochemistry (IHC)-based analyses of sequential biopsies in two xenograft models. These mimic phase 2 and 3 trials in our own therapeutic strategy to prevent the adaptive response to BRAF inhibition. In this mechanistic model, the dependencies between the targeted $Li_2CO_3$-inducible hub HuR and the resistance effectors are more likely time-shifted and transient since the minority of $HuR^{Low}$ cells, which act as a reservoir of adaptive plasticity, switch to a $HuR^{High}$ state as they paradoxically proliferate under BRAF inhibition. Nevertheless, we show that a copula/kernel density estimator (KDE)-based quantification of mutual information (MI) efficiently captures, at the individual level, the dependencies between HuR and two relevant resistance markers pERK and EGFR, and outperforms classic expression correlation coefficients. Ultimately, the validation of MI as a predictive IHC-based metric of response to our therapeutic strategy will be carried in clinical trials.

## I. INTRODUCTION

Based on a large amount of experimental research, our comprehension of the mechanisms of adaptive resistance to targeted therapies of solid mutated malignancies has significantly improved during the last decade [1]. Regardless of the current controversies on the scale-free properties of the cell/cancer signaling network [2], one of the most prevalent intuitions behind this experimental research has been to overcome the redundancy and robustness of such network by targeting its most essential connected nodes. Within the paradigmatic model of the adaptive response of *BRAF*-mutated melanoma to BRAF inhibition on which we focus here, the convergence on MYC activation of various upstream signaling pathways and downstream targets [3] or the targetable WNT5A-induced activation of the AKT pathway associated with transcriptional reprograming [4], stand as good examples of "hub-directed" strategies in the recent literature. Currently, one of the major challenges of such mechanistic studies is to confirm *in vivo*, at the single-cell proteomic level, the existing dependencies between these hubs and their connected nodes. Similarly, the drug-targeted *in vivo* modulation of these dependencies needs to be demonstrated. These *in vivo* analyses are often not performed because of the lack of sensitivity to change of the available quantification techniques and the non-linear and time-shifted mutual dependencies of the involved mechanistic factors.

Such quantification becomes even more challenging if the proportion of cells in which such dependency operates is small within the tumor tissue, particularly at the initial time-points of the adaptive response. This would be the case for rare, highly plastic cells that reprogram under therapeutic selection [5]. It would also occur in any embryonic signaling network that operates in the minority of senescent stem cell-like cells that give rise to adaptive resistance [6]. Nevertheless, these quantifications are necessary for any further clinical implementation of these hub-directed strategies, not only to confirm their involvement in the observed clinical outcome but also as to obtain predictive markers of response to these strategies.

Practically, these analyses would need to be performed using techniques that can be routinely performed on patient biopsies as immunohistochemistry (IHC). Here, to implement such single-cell based *in vivo* expression dependency quantification, we use mutual information and compare it to more commonly used approaches in our own mechanistic strategy to reduce the adaptive response to BRAF inhibition in melanoma. In our model, the targeted hub HuR/ELAVL1 (HuR) operates on the adaptive response in a minority of cells intermittently only if its expression becomes insufficient. Any dependency is therefore more likely time-shifted and transient. This model is therefore ideal to challenge the sensitivity of our approach.

* Corresponding author: rastine.merat@unige.ch
†These authors contributed equally to this work

## II. MATERIALS AND METHODS

### A. Descriptive statistics and mutual information estimation

All analyses, including tumor mean values Pearson's correlation, single-cell-based Spearman's correlation and mutual information estimation as well as the statistical tests, were conducted in MATLAB2020b (code for mutual information estimation and example dataset available upon request).

Mutual information (MI) is invariant to reparameterization, consequently, as a first step to decrease the impact of variability inherent to use of IHC, we use a copula-transform of values obtained for both markers (i.e., rank order them between 0 and 1) at the single-cell level. This initial step leads to a uniform distribution of their marginal distributions. We then estimate MI using a "smoothing" nonparametric Gaussian kernel density estimator (KDE), which is a local weighted average of the relative frequency of observations in the neighborhood of each estimate [7] given as

$$\hat{p}(x) = \frac{1}{N} \sum_{i=1}^{N} K(u), \qquad (1)$$

where $x$ is the two-dimensional signal-measured intensities for any single cell and $N$ is the number of samples (cells) and

$$u = \frac{(x - x_i)^T S^{-1} (x - x_i)}{h^2}, \qquad (2)$$

where $h$ is the bandwidth of the kernel smoothing window and $S$ is the covariance matrix of $x$. $K(u)$, the kernel function, is given as

$$K(u) = \frac{1}{(2\pi)^{d/2} h^d \det(S)^{1/2}} \exp(-u/2). \qquad (3)$$

Since $K(u)$ is a multivariate normal density function of dimension $d$, $h$ is calculated using the "optimal" Silverman's bandwidth that minimizes the mean integrated square error (MISE),

$$h = \left( \frac{4}{(d+2)N} \right)^{\frac{1}{d+4}}. \qquad (4)$$

Compared to classical histogram binning, this method is insensitive to the choice of origin, and most importantly provides a continuous better estimate of the underlying probability density, which avoids biases related to binning [7] or assumptions about the underlying distributions.

The one- and two-dimensional estimates obtained with this pipeline are then used to calculate the mutual information between the two variables $x, y$ (markers intensity) defined as

$$\hat{I}(x, y) = \frac{1}{N} \sum_{i,j} \log_2 \left( \frac{\hat{p}_{XY}(x_i, y_j)}{\hat{p}_X(x_i) \hat{p}_Y(y_j)} \right). \qquad (5)$$

To ensure the robustness of the MI estimates in each tumor and for each pair of markers, we randomly select the subpopulation of cells used to estimate the MI (one hundred assays per estimate and surrogate). The fixed number of cells used in this subpopulation across all biopsies being compared is defined as the smallest number of cells detected within all biopsies.

### B. Mouse xenografts

Animal experiments were approved by the Animal Welfare Commission of the Canton of Geneva (approval n° GE/108/18) and followed the Swiss guidelines for animal experimentation. For model 1, two million shCtrl or shHuR SK-MEL28 cells (FACS-determined 70% apparent shift), generated as previously described [8], were resuspended in 100 µl of PBS and mixed with an equal volume of Matrigel (Corning® Matrigel® Matrix High Concentration, Phenol-Red free) and injected subcutaneously into the posterior left flanks of six-week-old female immunodeficient NOD.Cg-Prkdcscid Il2rgtm1Wjl/SzJ mice (NSG, Charles River). Tumor formation was monitored twice a week using calliper measurements and calculated by the ellipsoidal formula: tumor volume = (length x width²) x 0.5. The vemurafenib was prepared using 240 mg vemurafenib tablets (for human use) that were manually ground and suspended in a water containing solution of carboxymethylcellulose (CMC, 1%) and DMSO (5%). The final concentration of vemurafenib was 16.5 mg/mL as confirmed using an in-house liquid chromatography-electrospray ionization-tandem mass spectrometry method. Once the tumors reached a volume of approximately 0.2 cm³, all mice were treated by oral gavage once a day with 150 µL of this suspension (fixed vemurafenib dosage of 100 mg/kg/day). All mice also received a lithium carbonate ($Li_2CO_3$) containing chow dosed at 0.25% (2.5g/kg). However, considering that in the initially conducted fast growing A375 cells experiment (model 2) [8], tumor growth was affected in the $Li_2CO_3$ arm and that in this experiment tumor growth was initially extremely slow in both shCtrl and shHuR arms, the $Li_2CO_3$ therapy was initiated at the regrowth time-point. This was mandatory to obtain large-enough tumors that were, upon mice sacrifice, immediately collected and formalin-fixed and subsequently paraffin-embedded (FFPE) (time-point 2, TP2, no time-point 1, TP1). The methodology for the A375 xenografts (model 2), comprising two groups receiving or not $Li_2CO_3$, has been previously reported [8]. In model 2, punch biopsies were collected under local anesthesia and similarly prepared (FFPE) in a separate-cohort of mice before treatment initiation (TP1$_{t0}$) and 10 days later (average volume-doubling time) in a subgroup of this cohort that did not receive any therapy during that time (TP1$_{t10}$). Samples were similarly prepared (FFPE) upon mice sacrifice at the end of the experiment (TP2).

### C. Immunohistochemistry (IHC)-based single-cell automated quantification

Standard fluorescence-based immunohistochemical staining on deparaffinized tumor sections was performed as previously described [8]. In short, 5-µm thick sections were deparaffinized and cleared using UltraClear™ reagent and rehydrated in ethanol. Following antigen retrieval in citrate buffer, tissues were permeabilized with 0,1% tween 20 and blocked in 5% normal goat serum (NGS). The mouse monoclonal anti-human HuR antibody 3A2 (1:100) was used for co-staining (2h exposure at room temperature in PBS tween 0,1% NGS 5%) with one of the following rabbit primary antibodies: anti-S100 A1 antibody (SAB502708, Sigma 1:100), anti-EGFR antibody (D38B1, Cell signaling, 1:50), anti-phospho-p44/42 MAPK (pERK1/2) antibody (520G11, Cell signaling, 1:200). Following washing, Alexa 488-conjugated anti-mouse, or Alexa 555-conjugated anti-rabbit antibodies (1:500) were used as secondary antibodies (1 h exposure at room temperature). Following additional washing, slides were mounted with Dapi FluoromountG (Southern Biotech). Images were acquired using an automated Zeiss Axioscan.Z1 with a 20x 0.8NA Plan Apochromat objective (lateral resolution 0.325 µm/pixel). Nuclei were segmented using QuPath v0.2.3 based on the Dapi fluorescent channel. All areas of homogeneous S100 staining available within the tumor section were initially defined as region of interest (ROI) and subsequently manually transposed for all other stainings. Each individual nucleus area was expanded of 5 µm to simulate a cytoplasmic area. Both areas were used to measure the nuclear and the whole-cell mean signal fluorescence intensity in each cell. HuR being mainly a nuclear protein, its signal was measured in the nuclear area, whereas EGFR, pERK, and S100 signals were measured in the whole-cell surface area.

Human biopsies obtained from patients with metastatic melanoma disease were similarly analyzed. Their use was approved by the Geneva ethics committee (study n°2017-01346). According to the Swiss Federal Law for research, a positive vote of an ethical committee in a retrospective study is sufficient to use patient data and materials for research purposes without further need of individual informed consents. All patient-related data were identified as previously described [8] at the University Hospital of Geneva and selected on the availability of the samples.

## III. RESULTS

The RNA-binding protein HuR has been extensively characterized as a ubiquitously expressed post-transcriptional orchestrator of differentiation [9], cell death [10] and an expression synchronizer of cell-cycle regulatory genes (11).

HuR has the ability to regulate many of the previously identified hubs within the adaptive network to BRAF inhibition and could potentially represent a "super-hub" within this network [3, 4, 12-14]. We have previously shown that a heterogeneous and intermittently lower expression of HuR, within a subpopulation of *BRAF*-mutated melanoma cells (HuR$^{Low}$ state), is induced upon their exposure to a BRAF inhibitor (BRAFi). This heterogeneous state is increasingly detected during the adaptive response and is not dependent on the proliferation status of the cell population. The HuR$^{Low}$ heterogeneous state in turn induces a heterogeneous and adaptive plastic expression of resistance markers and an adaptive response of the whole cell population to chronic BRAF inhibition. Experimentally, in order to increase the adaptive response, the insufficient expression of HuR needs to be reversible between two attractor sets (reversible knockdown). Indeed, a stable knockdown of HuR has no effect on the adaptive response [8]. This observation indirectly indicates that although the heterogeneous HuR$^{Low}$ cells are a reservoir of adaptive plasticity, they need to switch to a HuR$^{High}$ state in order to proliferate. At the single-cell and at steady state, although the adapted highly proliferating cells are in a HuR$^{High}$ state and have the highest expression level of resistance markers, their emergence occurs within the HuR$^{Low}$ cell subpopulation carrying the heterogeneous HuR$^{Low}$ cell component. From a therapeutic standpoint, a slight lithium salt-induced suppression of the heterogeneous HuR$^{Low}$ cell component attenuates the adaptive paradoxical expression and proliferative response to BRAF inhibition [8].

To develop *in vivo* predictive expression dependency markers of response to the resulting therapeutic strategy of combining lithium salts with small-molecule inhibitors in metastatic *BRAF*-mutated melanoma, we conduct here a series of analyses on biopsies obtained from two mouse xenograft models established according to the design of a phase 2 (model 1) and phase 3 (model 2) clinical trials in which biopsy samples would be obtained for translational studies. We focus our IHC-based monitoring on the expression of HuR and its dependent resistance markers, EGFR and pERK. *Ex vivo*, the increased expression of EGFR or pERK under reversible knockdown of HuR or conversely, their decreased expression under suppression of the HuR$^{Low}$ cells by lithium salts, are clearly detected at the whole cell population level in synchronized cells adapted to, and treated with, a BRAF inhibitor [8]. However, at steady state and under physiologic variation of HuR expression, these changes are difficult to detect, considering that they occur at any moment in a subpopulation of adapting cells. Nevertheless, at the single-cell level, the dependencies between HuR and these resistance markers are expected to change in adapting cells and be detectable *in vivo*. Conversely, our results predict that these changes will be modulated in tumors concomitantly exposed to lithium salts.

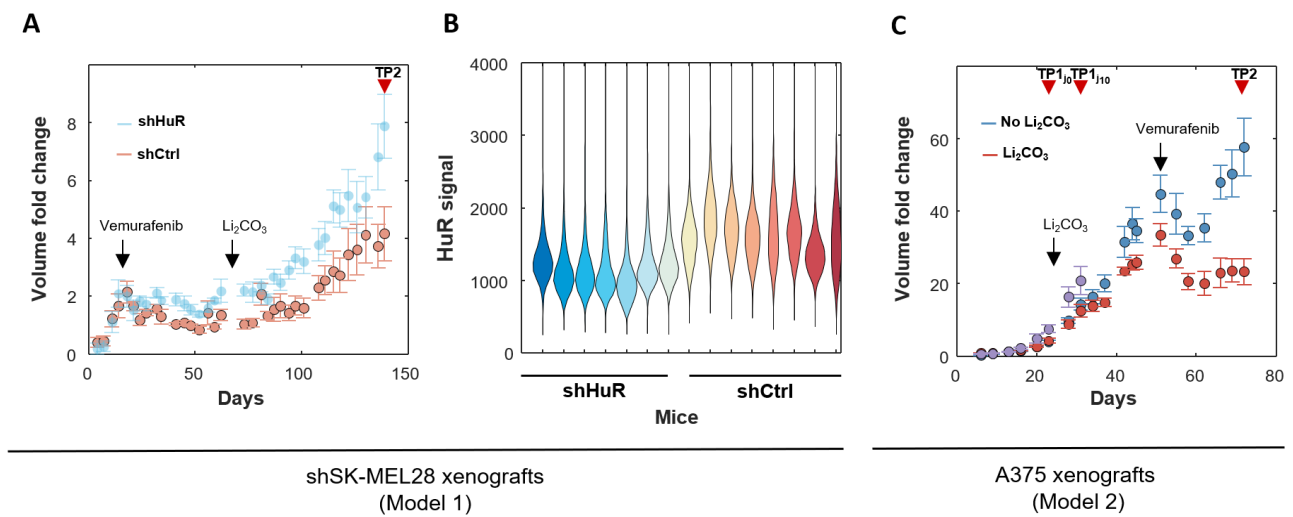

FIG. 1. Mice xenografts melanoma models of adaptive response to BRAF inhibition used in this study. (A) Model 1: SK-MEL28 *BRAF*-mutated (i) shHuR (HuR lithium-non inducible) carrying an unstable proportion of HuR$^{Low}$ cells oscillating between two attractor sets [8] (blue) (n=7), (ii) shCtrl (HuR lithium-inducible) (red) (n=8). Mice were treated as in a single-arm phase 2 trial and received the BRAFi vemurafenib (100mg/kg/day) with subsequent addition of lithium carbonate (Li$_2$CO$_3$) containing chow dosed at 0.25% (2.5g/kg) at the indicated time-points. Considering that before exposure to the vemurafenib, HuR heterogeneous expression was already operating at a much higher extent in the shHuR panel than the one expected to be induced by the treatment, excision-biopsies were performed only at the end of the experiment (TP2, no TP1). (B) IHC-based HuR expression distribution in shHuR (blue) and shCtrl (red) xenografts shown as violin plots. (C) Model 2: A375 *BRAF*-mutated xenografts. Mice were treated as in a double-arm phase 3 trial and assigned to control (blue) or Li$_2$CO$_3$ chow (red) dosed as in (A) and subsequently all received the BRAFi vemurafenib dosed as in (A) at the indicated time-points. Biopsies were performed before initiating the Li$_2$CO$_3$ therapy (TP1$_{j0}$, n=15) and following an average volume-doubling time period of 10 days in a subcohort of mice that did not receive any therapy during that time (TP1$_{j10}$, n=7, purple). Final excision-biopsies were performed in both arms at the end of the experiment (TP2 and TP2$_{Li_2CO_3}$). For (A) and (C) data shown are mean tumor volume (indicated as fold change) ± SEM.

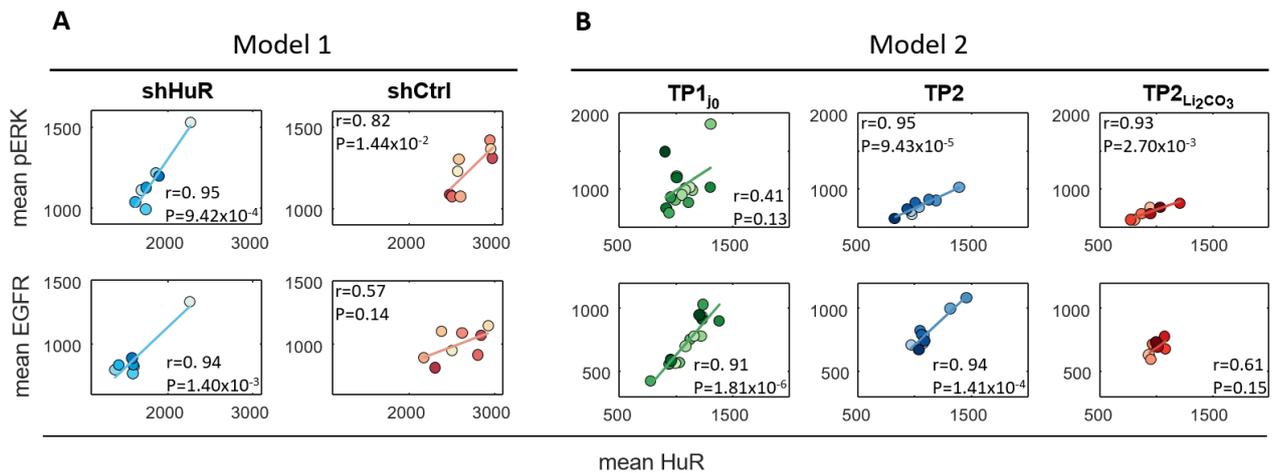

FIG. 2. HuR/pERK and HuR/EGFR linear dependencies examined based on the strength of Pearson's correlation between the average expression value of markers in tumor biopsies performed, (A) in model 1 in SK-MEL28 shHuR (blue) and shCtrl (red) xenografts, (B) in model 2 in A375 cells xenografts before initiating any therapy (TP1$_{j0}$, green) and at the end of the experiment in the vemurafenib + control chow arm (TP2, blue) or the vemurafenib + Li$_2$CO$_3$ chow arm (TP2$_{Li_2CO_3}$, red) (see Fig. 1 legend). In each panel, the Pearson's correlation r and p-value for testing the hypothesis of no correlation is indicated.

In model 1 (Fig. 1A), designed as a single-arm phase 2 trial, all individuals received the BRAFi vemurafenib with subsequent addition of lithium carbonate (Li$_2$CO$_3$) to their regimen. The SK-MEL28 HuR reversible knockdown cells (shHuR) carrying an unstable proportion of HuR$^{Low}$ cells [8], were used to generate a panel of xenografts (n=7) in which the average expression of HuR is variable yet distinguishable from, and inferior to, the average expression of HuR observed in the control (shCtrl) panel (n=8) (Fig. 1B). Based on our previous *ex vivo*-made sensitive immunocytochemistry analyses, HuR expression is not inducible in these cells as opposed to their shCtrl counterparts in which a positive shift in HuR nuclear and cytoplasmic content is detected upon exposure to therapeutic concentrations of Li$_2$CO$_3$ [8]. This *in vivo* experimental model is therefore perfectly suited to measure to what extent the dependencies between HuR and both EGFR and pERK are detectable and distinguishable between non-responders' tumors carrying a more heterogeneous expression of HuR and the responders' carrying a more homogeneous expression of HuR. In this model, before treatment initiation, a HuR heterogeneous expression is already operating to a much higher extent in the shHuR panel than the one expected to be induced by the BRAFi; the comparative analyses were therefore performed only on biopsies obtained at the end of the experiment, from tumors exposed to sustained BRAF inhibition (time-point 2, TP2, no time-point 1, TP1).

In model 2 (Fig. 1C), designed as a double-arm phase 3 trial, individuals (n=16) were randomly assigned into two arms to receive or not Li$_2$CO$_3$ in addition to the vemurafenib treatment. The A375 *BRAF*-mutated melanoma cells were used to generate the xenografts. These cells have a high deterministic behavior for simulating tumor relapse following an initial response to BRAF inhibition and are therefore often used as a model of adaptive response to BRAF inhibition [15]. Our *ex vivo* observations indicate that HuR expression is more heterogeneous in A375 cells having a higher propensity for adaptive response than in cell lines having a lower propensity for adaptive response, e.g., the SK-MEL28 cells used in model 1. Moreover, this expression heterogeneity is highly increasing in A375 cells following exposure to a BRAFi [8]. Therefore, contrary to model 1, a change in the baseline expression dependencies between HuR and both EGFR and pERK is expected to occur in BRAFi-treated tumors. For this reason, the comparative analyses were performed on biopsies obtained both before treatment initiation (n=15, separate cohort, TP1$_{j0}$, see materials & methods) and upon tumor regrowth (TP2). Moreover, the tumor volume is rapidly increasing in this model and might affect the expression dependencies of the markers. Consequently, additional biopsies were performed following an average volume-doubling time period of 10 days in a sub-cohort of individuals (n=7) that did not receive any treatment during that time (TP1$_{j10}$). The overall outcome of this experiment was previously reported and showed that following the initial response to vemurafenib, relative tumor regrowth was clearly attenuated in the Li$_2$CO$_3$ receiving arm (Fig. 1C) [8].

## A. Average expression correlations

HuR/pERK and HuR/EGFR dependencies were first examined based on the strength of Pearson's correlation between the tumor average expression values of markers (Fig. 2). In model 1 (Fig. 2A), an overall positive correlation for HuR/pERK and HuR/EGFR was apparent and consistent with our *ex vivo* single-cell observations, in which the adapted HuR[High] proliferating cells showed the highest expression level of resistance markers [8]. Nevertheless, for HuR/pERK, the strength of correlation was higher in the shHuR panel, and for HuR/EGFR, a positive correlation was only observed in the shHuR panel. In model 2 (Fig. 2B), HuR/EGFR already positively correlated at baseline in TP1$_{j0}$ biopsies and a HuR/pERK correlation was induced following exposure to BRAFi in TP2 tumors, compared with baseline. However, the suppression of these correlations/dependencies in TP2$_{Li_2CO_3}$ tumors was only apparent for HuR/EGFR. Overall, these results were consistent with the mechanistic model deduced from our *ex vivo* experimental results in which the dependency of resistance markers toward HuR increases when HuR expression is rendered insufficient: whether experimentally as in model 1 reversible knockdown, or therapeutically following exposure to a BRAFi, as for HuR/pERK in model 2. Yet, the Li$_2$CO$_3$-induced suppression of these dependencies in model 2 was only captured for HuR/EGFR. Moreover, the overall sensitivity of this commonly used approach was undoubtedly insufficient to predict the phenotypic outcome of individual tumors i.e., in terms of predicting in model 2 the therapeutic arm to which they belong.

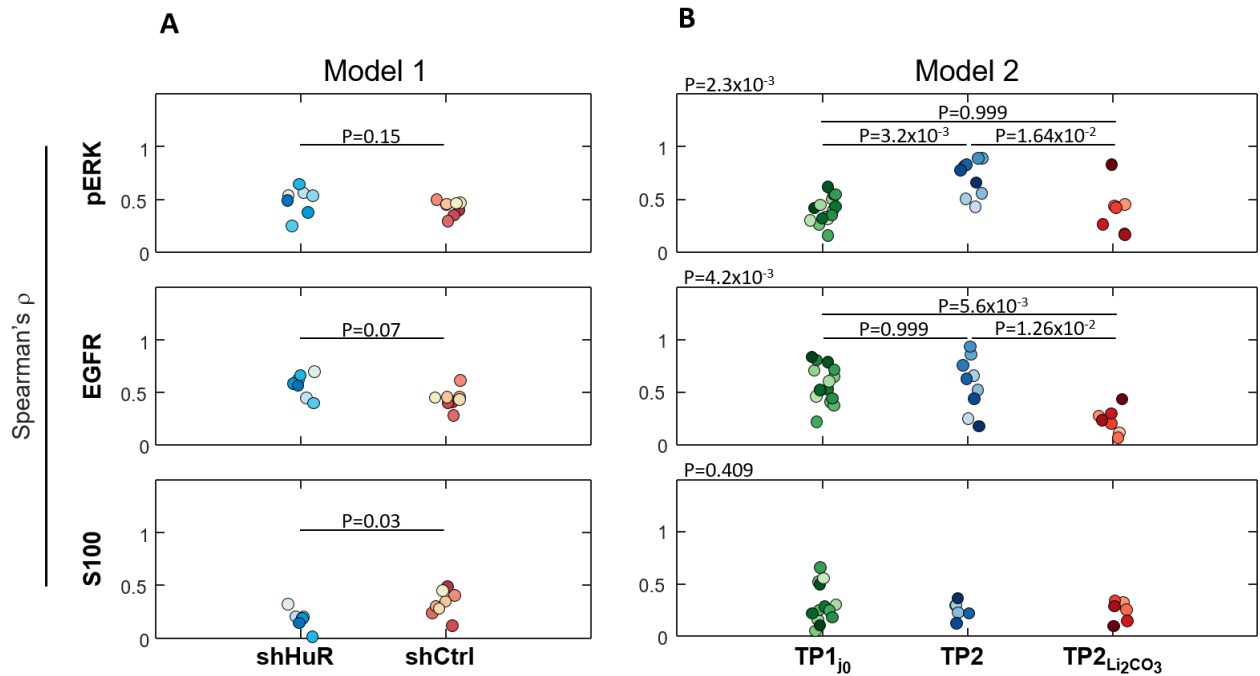

FIG. 3. HuR/pERK, HuR/EGFR and HuR/S100 dependencies examined based on the Spearman's rank correlation ρ value of expression of the markers at the single-cell level in tumor biopsies performed as in Fig. 2. For the two models, Fisher transformations were applied on each correlation. In model 1, Wilcoxon rank sum tests were used to compare medians. In model 2, Kruskal-Wallis tests were used to compare the distributions (upper left p-value), when the null hypothesis was rejected post-hoc Tukey-Kramer tests were performed.

## B. Single-cell Spearman's correlations

Next, in order to improve our sensitivity to detect non-linear dependencies, we calculated for each sample the Spearman's (rank) correlation ρ for HuR/pERK and HuR/EGFR at the single-cell level (Fig. 3). Henceforward, the expression of the differentiation marker S100 was also included as a "negative" marker of dependency. Indeed, the dependency between HuR and S100 is expected to either be low in the undifferentiated cell lines used in both models or at least not to be positively selected in the adaptive response to BRAF inhibition. In model 1 (Fig. 3A), strikingly, no significant differences in the ρ values were observed between the shHuR and the control panels. In model 2 (Fig. 3B), consistent with the average expression correlations analyses, the ρ coefficient was significantly higher for HuR/pERK upon exposure to the BRAFi in TP2 tumors compared with the baseline TP1$_{j0}$ biopsies. Compared with TP2 tumors, here, a significant reduction of ρ was observed in TP2$_{Li_2CO_3}$ tumors for both HuR/pERK and HuR/EGFR dependencies but was more apparent for the latter. For HuR/S100, as expected no significant differences were observed for ρ values among the samples. These analyses were therefore confirmatory and more sensitive for detecting the Li$_2$CO$_3$-induced suppression of both HuR/pERK and HuR/EGFR dependencies in model 2. However again, they were insufficient in predicting the phenotypic outcome of individual tumors and even in detecting the differences in dependencies in model 1. In this model, the lack of sensitivity was presumably due to an averaging effect between HuR[Low] and HuR[High] cells, both types being, according to our *ex vivo* experimental results, timely involved in these dependencies.

## C. Mutual information

MI is a universal metric for quantifying any type rather than just linear dependencies between two variables. Information theory and entropy-based description of MI provide the formalism to define MI *I* between two variables *X* and *Y* as

$$I(X;Y) = H(X) + H(Y) - H(X,Y), \qquad (6)$$

where *H* is the entropy associated with each of the variables, i.e. the amount of information gained (or reduced uncertainty) on each of them when measuring them separately, and $H(X,Y)$ is the joint entropy symmetrically associated with both variables, i.e. the amount of information gained on one of the variables when measuring the other variable. Most importantly, MI is scale-free and its quantification using rank-ordered data (see materials and methods) allows to overcome some of the limitations of semi-quantitative techniques as IHC [16, 17]. These include variability in sample preparation and imaging. However, as extensively discussed in the physics literature, MI is highly dependent on the size of the probability distribution and is sample-size-dependent [18]. This is even true for discrete data for which, in order to get a reliable estimate of MI, the number of samples needs to be significantly larger than the cardinality of the underlying distribution. To overcome these limitations, we used a computationally efficient Gaussian kernel estimator (KDE see materials and methods) and applied it throughout this study on a fixed smallest number of cells detected within the samples being compared. With this fixed number, cells were randomly chosen in the one hundred assays which were performed to ensure the stability of each MI estimate.

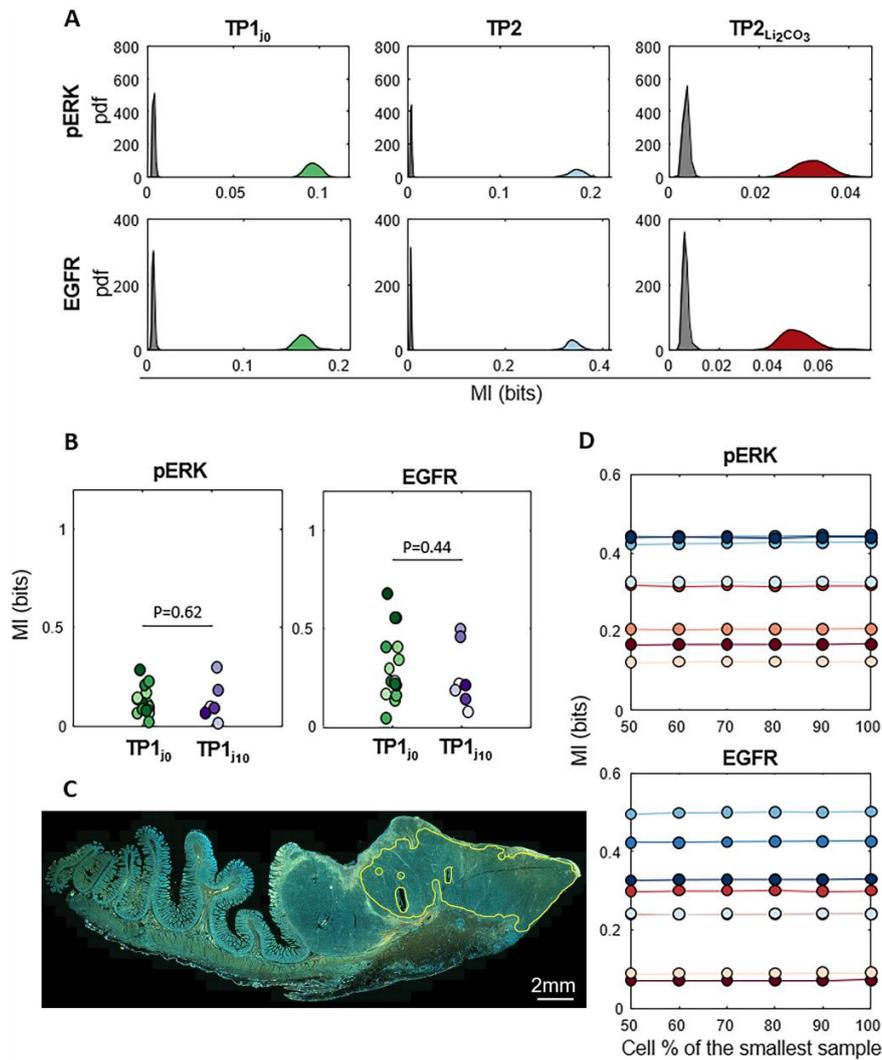

FIG. 4. Robustness of the MI estimates including in respect to the probability distribution size and the sample size biases. (A) Examples of HuR/pERK and HuR/EGFR MI estimates in the three types of biopsies obtained in model 2. To calculate surrogates, one hundred assays were conducted by randomly shuffling the intensity of the second marker intensity in the subpopulation of cells used for MI estimate (grey histogram). (B) Tumor growth effect on HuR/pERK and HuR/EGFR MI estimates. Biopsies from baseline (TP1$_{j0}$, green) are compared with identical tumors following a volume-doubling time of 10 days in a subcohort of mice that did not receive any therapy during that time (TP1$_{j10}$, purple). Testing the hypothesis of equal MI under tumor growth was done by using Wilcoxon rank sum tests. (C, D) Sample size effect on HuR/pERK and HuR/EGFR MI estimates in human metastatic melanoma. (C) Example of HuR/pERK co-staining performed on a small intestine metastatic melanoma disease. Note that the region of interest (ROI) was chosen as to be similar to the additional tumor sections that were used for other stainings. (D) The fixed number of cells (percentage of the smallest sample) being used for HuR/pERK and HuR/EGFR MI estimates is changed across eight biopsies of metastatic disease. Biopsies were performed before small-molecule inhibitor therapies were initiated but only half of the patients had a complete response (cream, red and brown color range, n=4), the remaining patients had either partial response or progressive disease (blue color range, n=4) on the biopsied tumor two to three months after treatment initiation (radiologically-assessed response according to RECIST1.1).

We first demonstrated that in each tumor sample and for each pair of markers, the MI estimates were distinguishable from a "null" MI obtained by shuffling one of the two signals across cells (Fig. 4A). An above null MI was quantifiable even for samples having a very low HuR/pERK or HuR/EGFR MI estimate e.g., the TP2$_{Li_2CO_3}$ tumors in model 2. Although using a KDE, we then asked if the size of the probability distribution had an impact on the MI estimates. We considered that a natural increase in the probability distribution might occur during tumor growth even if the fixed-size sample is randomly chosen in the tumor section, and estimated the HuR/pERK and HuR/EGFR MI of biopsies obtained in a subcohort of mice that did not receive any treatment during an average volume-doubling time period of 10 days (TP1$_{j10}$). These MI estimates were not significantly different from the one obtained in biopsies performed on smaller TP1$_{j0}$ identical tumors (Fig. 1C and 4B). Next, we checked if the sample size would affect the MI estimate by varying the number of cells analyzed within tumors. Considering that the xenografts in our mouse models were likely homogeneous in their cell composition, these analyses were carried on biopsies (n=8) obtained from patients with *BRAF*-mutated metastatic melanoma disease before targeted therapy was initiated. Once treated, these patients had various types of response in their biopsied tumors which were therefore more likely heterogeneous in their cell composition. Yet the change in sample size across all metastatic tumors had no effect in their MI estimates and ranking (Fig. 4C and D). Overall, we concluded that our copula-transformed/KDE-based estimate of MI was robust enough to overcome both the probability distribution size and the sample size biases.

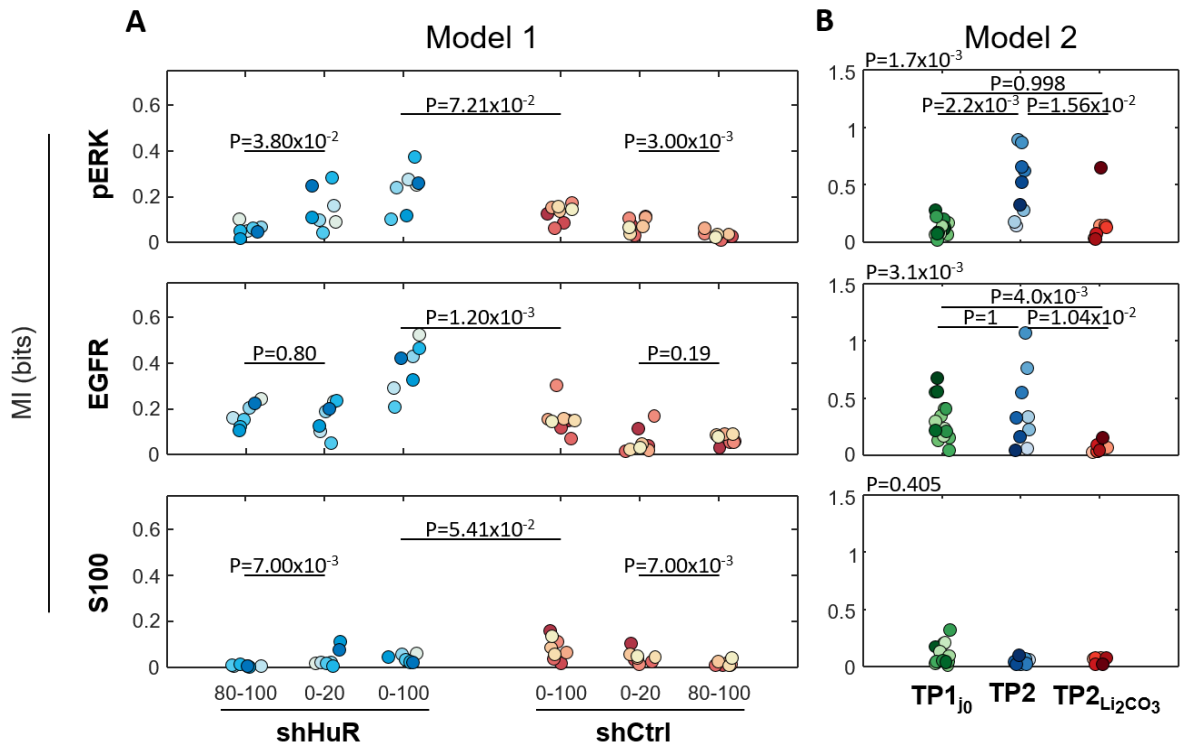

FIG. 5. HuR/pERK, HuR/EGFR and HuR/S100 dependencies examined based on MI estimate as in Fig. 2. For model 1, additional estimates of MI were calculated for the 20th percentile of the highest HuR-expressing cells (80-100) and for the 20th percentile of the lowest HuR-expressing cells (0-20). In model 1, Wilcoxon rank sum tests were used to compare medians. In model 2, Kruskal-Wallis tests were used to compare the distributions (upper left p-value), when the null hypothesis was rejected post-hoc Tukey-Kramer tests were performed.

In model 1, MI estimates for both HuR/pERK and HuR/EGFR were significantly higher in the shHuR tumors than in the control panel (Fig. 5A). In contrast, HuR/S100 MI, used here as a control, was slightly but significantly higher in the control panel. Remarkably, HuR/EGFR MI estimates were almost sufficient to individually distinguish the shHuR from the shCtrl tumors. In addition, contrary to the HuR/pERK MI estimates for the 20th percentile of the highest HuR-expressing cells, these estimates for the 20th lowest percentile were almost as high as the ones obtained for the total cell population in most shHuR tumors, suggesting that most of the HuR/pERK dependency was operating in the HuR$^{Low}$ cells in these tumors. Overall, the increased HuR/pERK and HuR/EGFR MI-based dependencies observed in shHuR tumors were consistent with the average expression Pearson's coefficient-based analyses discussed in section 3.1, however, MI performed better, particularly in capturing HuR/EGFR dependency in individual tumors.

In model 2 (Fig. 5B), in accordance with 3.1 results, HuR/EGFR MI-determined dependency was already detected at baseline in TP1$_{j0}$ biopsies and became significant compared to baseline for HuR/pERK in adapted TP2 tumors. As expected, HuR/S100 MI estimate was low and not affected in BRAFi-treated tumors. Importantly, the Li$_2$CO$_3$ suppression of HuR/EGFR and BRAFi-induced HuR/pERK dependencies, under physiologic expression levels of HuR, was clearly captured here. With the exception of a few mice, MI performed well to distinguish the mice co-treated with Li$_2$CO$_3$ from the ones treated only with the BRAFi.

Taken together, these results support the use of MI, rather than the classical metrics used in sections 3.1 and 3.2, for quantifying HuR/pERK and HuR/EGFR dependencies in our therapeutic strategy. The increase in these dependencies under an experimentally induced insufficient expression of HuR occurred similarly, at least for one pair of markers, under BRAFi therapy and were reduced under concomitant Li$_2$CO$_3$ therapy. More importantly, the predictive value of MI at the individual level was far more effective. Ultimately, the validation of MI as a biologic, predictive IHC-based metric of response to combining small-molecule inhibitors with Li$_2$CO$_3$ in *BRAF*-mutated metastatic melanoma will be carried in trials equivalent to either one of the models developed here.

## IV. DISCUSSION

Although extensively used in the literature to elucidate gene regulatory networks [17,19] and quantify cellular signal processing [20, 21], MI-based quantification of expression dependencies has not been used, to our knowledge, as an *in vivo* confirmatory approach for mechanistic studies, nor as a quantitative metric for clinical use. As shown in this work, MI is far more sensitive to change and performs better to predict drug effect than average co-expression or even single-cell-based co-expression correlation coefficients. MI should be particularly used when suspecting complex patterns of dependencies and when dealing with dynamic, reversible dependencies as observed during the adaptive response to targeted therapies in solid malignancies. This approach may help clarify contradictory observations made in this field. As an example, the melanocytic lineage transcription factor MITF has been reported to be either upregulated [22] or downregulated [23, 24] in association with an increased level of pERK in melanoma cells adapted to and treated with a MAPK inhibitor. These contradictory observations could either be attributed to the use of different experimental settings including cell types or be related to partial independencies between these markers (in which case the presumed mechanistic effect is at least partially refuted). However, they could as well be related to a heterogeneous, complex, pattern of dependency [25] that could be captured by MI as the one described in this study between HuR and EGFR or pERK.

Generally, the *in vivo* dynamic change in MI between the main regulatory hub and its targets should be first assessed in an experimentally-induced, presumed pathologic expression condition of the targeted hub and subsequently, under its physiologic expression where the incriminated pathologic change of MI is expected to be therapeutically reversed or at least modulated according to the observed phenotypic outcome. Importantly, the integration of the methodology developed here into the translational analytic pipeline used to develop predictive markers of response to therapy in clinical trials is highly feasible. Its potential future use in routine pathology as an automated procedure is therefore predicted.

## ACKNOWLEDGMENTS

We thank Wolf-Henning Boehncke for continuous support, Ludovic Wrobel for his assistance for performing the xenografts and animal biopsies, Youssef Daali for conducting the vemurafenib dosage in the gavage solution, Pierre Bonnaventure and his team for performing the mice gavage.

This work was supported by the Ligue Genevoise contre le Cancer and the Fondation pour la lutte contre le cancer (Zurich).

## AUTHOR CONTRIBUTIONS

Aurore Bugi-Marteyn performed the animal experiments and the immunohistochemistry procedures. Fanny Noulet performed all the confirmatory immunohistochemistry procedures. Nicolas Liaudet designed the code, analyzed the data and designed the figures. Rastine Merat designed the theoretical framework, supervised the project, analyzed the data, designed the figures and wrote the manuscript.

## DECLARATION OF INTEREST

Rastine Merat is inventor on a patent on the use of agents enhancing HuR/ELAV protein levels in the treatment of BRAF-mutated cancers. Declarations of interest for Aurore Bugi-Marteyn, Fanny Noulet and Nicolas Liaudet: none.

---